\documentclass[iop,apj]{emulateapj}
\usepackage{cases}
\usepackage{graphicx}

\begin{document}

\title{Cosmic ray acceleration at perpendicular shocks \\in supernova remnants} 

\submitted{accepted for publication in ApJ July 23, 2014}

\author{Gilles Ferrand$^1$, Rebecca J. Danos$^1$, Andreas Shalchi$^1$, Samar Safi-Harb$^1$\altaffilmark{,2}, Paul Edmon$^3$, Peter Mendygral$^{4,5,6}$}
\affil{${}^1$Department of Physics and Astronomy, University of Manitoba, 
Winnipeg, MB, R3T 2N2, Canada\\${}^3$Harvard University, 60 Garden Street, Cambridge, MA 02138, USA \\${}^4$School of Physics and Astronomy, University of Minnesota,
Minneapolis, MN 55455, USA\\${}^5$ Minnesota Supercomputing Institute, University of Minnesota, Minneapolis, MN 55455, USA\\${}^6$Cray, Inc., 380 Jackson Street, Suite 210, St. Paul, MN 55101, USA}
\email{gferrand@physics.umanitoba.ca}
\altaffiltext{2}{Canada Research Chair}

\begin{abstract}
Supernova remnants (SNRs) are believed to accelerate particles up to high energies through the mechanism of diffusive shock acceleration (DSA). Except for direct plasma simulations, all modeling efforts must rely on a given form of the diffusion coefficient, a key parameter that embodies the interactions of energetic charged particles with the magnetic turbulence. The so-called Bohm limit is commonly employed. In this paper we revisit the question of acceleration at perpendicular shocks, by employing a realistic model of perpendicular diffusion. Our coefficient reduces to a power-law in momentum for low momenta (of index~$\alpha$), but becomes independent of the particle momentum at high momenta (reaching a constant value $\kappa_{\infty}$ above some characteristic momentum~$p_{\rm c}$). We first provide simple analytical expressions of the maximum momentum that can be reached at a given time with this coefficient. Then we perform time-dependent numerical simulations to investigate the shape of the particle distribution that can be obtained when the particle pressure back-reacts on the flow. We observe that, for a given index~$\alpha$ and injection level, the shock modifications are similar for different possible values of $p_{\rm c}$, whereas the particle spectra differ markedly. Of particular interest, low values of $p_{\rm c}$ tend to remove the concavity once thought to be typical of non-linear DSA, and result in steep spectra, as required by recent high-energy observations of Galactic SNRs.
\end{abstract}


\maketitle
\newpage


\section{Introduction}
\label{sec:introduction}

Cosmic rays (CRs) observed on Earth possess a wide range of energies, with a spectral break (the ``knee") at approximately $10^{15}$ eV. Particles accelerated up to this energy are thought to originate in Galactic sources, such as supernova remnants (SNRs). 
The observation of non-thermal high-energy radiation from SNRs is amongst the accumulating evidence that these objects are indeed accelerators of particles (for a recent review see \citealt{2012SSRv..173..369H}). The observation of X-ray synchrotron rims in young SNRs proved the presence of multi-TeV electrons (first in SN~1006, \citealt{1995Natur.378..255K}), and the detection of the $\gamma$-ray pion-decay signature in more evolved objects recently confirmed the presence of energetic protons as well (in IC~443 and W44, \citealt{2013Sci...339..807A}).  
It is generally believed that particles achieve such high energies via the mechanism known as diffusive shock acceleration (DSA), a first-order Fermi acceleration process \citep{1978MNRAS.182..147B,1983RPPh...46..973D, 1987PhR...154....1B, 2001RPPh...64..429M}. In DSA, particles increase their energy by crossing the shock front multiple times, scattering off turbulence in the magnetic field. The main prediction of DSA is that the final CR distribution function is a power-law function in momentum: $f(p) \propto p^{-3r/(r-1)}$ where $r$ is the shock compression ratio. For strong shocks ($r=4$) this results in an index of~4, globally consistent with observations \citep[e.g.][]{2011Ap&SS.336..263K}.

In order to describe the mechanism of DSA, one has to solve the CR transport equation, which is essentially an extended diffusion equation \citep[see e.g.][]{2002cra..book.....S}. In general this equation contains a diffusion tensor describing the particle diffusion in the different directions of space. In the scenario described in the present paper, we are interested only in particle diffusion along the direction of the shock propagation, and therefore consider one-dimensional diffusion. So we solve the following form of the transport equation \citep[][]{1975MNRAS.172..557S}:
\begin{equation}
\frac{\partial f}{\partial t} + u\frac{\partial f}{\partial x} 
= \frac{\partial}{\partial x} \left(\kappa \frac{\partial f}{\partial x}\right)
+ \frac{1}{3}\left(\frac{\partial u}{\partial x}\right) \frac{\partial f}{\partial \ln p} 
+ S.
\label{eq:transport}
\end{equation}
where $f(x,p,t)$ is the CR distribution function of position~$x$ along the shock normal, momentum~$p$, and time~$t$, $u$ is the bulk flow speed, $\kappa$ is the spatial diffusion coefficient in the shock propagation direction, and $S$ is a source term. In this form of the transport equation we have neglected stochastic acceleration and other effects such as radiative losses. 

The key parameter $\kappa$ embodies the interaction between charged particles and magnetic turbulence. Computing its value is a difficult problem in theoretical astrophysics (see \citealt{2009ASSL..362.....S} for a review). In studies of DSA at interstellar shocks, the so-called Bohm limit is often employed (see, e.g., \citet{1991MNRAS.249..439K,1997APh.....7..183B,2013MNRAS.431..415B}). It is often assumed that the parallel diffusion coefficient cannot be smaller than the Bohm limit, and that the perpendicular diffusion coefficient cannot be larger than the Bohm limit. An alternative, simple model is to assume a power-law in momentum. The Bohm coefficient gets proportional to~$p$ for large~$p$, and so a common choice is a power-law of index $\alpha=1$, often referred to as ``Bohm-like" diffusion (see, e.g., \citet{1991MNRAS.249..439K,1997ApJ...485..638M,2002A&A...395..943B,2006APh....25..246K, 2010JKAS...43...25K,2012ApJ...745..146K}). Such a~steep dependence of the diffusion coefficient on momentum can make it very time-consuming to solve the transport equation computationally, because of the wide range of diffusion length- and time-scales to resolve. As a result, values of~$\alpha$ lower than unity have often been used for convenience. For instance, the early numerical simulations by \cite{1987MNRAS.225..399F} and \cite{1987MNRAS.225..615B} were made with respectively $\alpha=0.25$ and $\alpha=0.5$. Later on \cite{1991MNRAS.249..439K} performed a comparison of various indices between~0 and~1. At this point it should be noted that the choice of Bohm diffusion was historically favoured mostly because it provides the fastest possible acceleration at parallel shocks, for a given magnetic field. Although there is some observational evidence in favour of this scenario \citep{2006NatPh...2..614S,2007Natur.449..576U,2008ApJ...685..988T}, numerical studies have found that the Bohm regime does not generally hold \citep{2002PhRvD..65b3002C,2004JCAP...10..007C,2004NuPhS.136..169P}, and the validity of this assumption has been regularly questioned \citep{2001JPhG...27.1589K,2006A&A...453..387P}. To the best of our knowledge, the Bohm limit has been rigorously derived only in the presence of strong turbulence \citep{2009APh....31..237S,2014Ap&SS.350..197S}. We note that in studies of DSA in other contexts, such as interplanetary shocks \citep[e.g.][]{2010AdSpR..46.1208D,2012AdSpR..49.1067L}, different, more realistic diffusion models are used. 

It is the purpose of this paper to explore how the diffusion model impacts the spectrum of CRs produced through DSA at SNR shocks. Most of the literature cited assumes that SNR shocks are quasi-parallel shocks, where particles diffuse along the main magnetic field lines. However it is well known that perpendicular shocks, where particles diffuse across the magnetic field lines, work in a different (and more complicated) manner. It had been suggested some time ago \citep{1987ApJ...313..842J} that perpendicular shocks could actually be faster accelerators, including in SNRs. This idea has been sporadically used \citep[e.g.][]{2006A&A...454..687M}, but may not have received the attention it deserves. In this paper we revisit the problem of acceleration by perpendicular shocks in the context of recent advances in the theory of perpendicular diffusion. Both analytical work \citep{2010Ap&SS.325...99S} and numerical work \citep{2012AdSpR..49.1643Q} suggest that a distinctively different form of the diffusion coefficient may be applicable for perpendicular shocks. \citet{2013ApJ...764...37B} demonstrated that this model agrees with observations for the starburst galaxy NGC~253. In this model, the diffusion coefficient increases with momentum until a characteristic momentum is reached and then asymptotes. In this paper we explore the impact of this alternative model in the context of SNRs, and compare it with more conventional models. 

The outline of the paper is as follows. In section~\ref{sec:models}, we present the different models for the diffusion coefficient in detail. In section~\ref{sec:ana} we investigate analytically, in the linear regime, the maximum momentum that can be reached by particles. In section~\ref{sec:num} we investigate numerically, in the non-linear regime, the shape of the spectrum of accelerated particles. Simulations are made with the code {\sc Marcos} \citep{2008MNRAS.383...41F}, with parameters based on \citet{2009ApJ...695.1273K} (hereafter KRJ09). In section~\ref{sec:conclusion} we give our conclusions and discuss future works.


\section{Models for the Diffusion Coefficient}
\label{sec:models}

As already noted, in the general case, diffusion of energetic particles is a difficult topic due to the complicated interaction between particles and magnetic fields. The magnetic field configuration can be written as $\vec{B} = \vec{B}_0 + \delta \vec{B}$ where $\vec{B}$ is the total magnetic field, $\vec{B}_0$ is the mean magnetic field, and the component $\delta \vec{B}$ describes the turbulence. The latter field causes diffusion of particles in different directions with respect to the mean field. A~strong simplification of the diffusion tensor can be achieved if we assume axi-symmetry with respect to $\vec{B}_0$. In this case only two spatial diffusion parameters enter the transport equation, namely the parallel diffusion coefficient $\kappa_{\parallel}$ and the perpendicular diffusion coefficient $\kappa_{\perp}$. Usually these diffusion parameters are very different because different physical processes are leading to the transport of particles along and across the mean field. In this work we solve the one-dimensional diffusion Equation~(\ref{eq:transport}), which contains a single diffusion coefficient~$\kappa$, the diffusion coefficient in the direction of the shock propagation (along the shock normal). In the following we discuss two existing models for the parallel diffusion coefficient and one third new model for the perpendicular diffusion coefficient. The first two models can be employed if a parallel shock is assumed (for which $\kappa=\kappa_{\parallel}$) and the third one for perpendicular shocks (for which $\kappa=\kappa_{\perp}$). 

\subsection{The Bohm Limit}
\label{sec:models:Bohm}

The simplest form that can be used to model the spatial diffusion of particles is that of isotropic Bohm diffusion. In this particular model it is assumed that for a strong turbulent magnetic field $\delta B \gg B_0$, scattering of particles becomes isotropic and the corresponding mean free paths are equal to the Larmor radius $\lambda=R_L$. From the Newton-Lorentz equation the latter is given (in Gaussian units) by $R_L = p c /(q B_0)$, where $p$ is the particle momentum, $c$ is the speed of light, and $q$ is the absolute value of the particle charge. Of course the question arises as to what the Larmor radius is in a partially turbulent magnetic system. If the turbulence were not present, the particle's orbit would be a perfect helix with a well-defined radius~$R_L$. If there is strong turbulence, however, the particle's motion is more of a chaotic motion, where the propagation direction is changing rapidly in time, and the Larmor radius is no longer defined. It was shown by \cite{2009APh....31..237S} and later by \cite{2014Ap&SS.350..197S} that in the strong turbulence limit one can find for the parallel mean free path $\lambda_{\parallel} \approx R_L\:B_0 / \delta B$ where $R_L$ is the unperturbed Larmor radius. The latter formula is different compared to the standard Bohm model (where $\lambda=R_L$) and it was confirmed recently by using test-particle simulations \citep[see][]{2014ApJ...785...31H}. Particle diffusion coefficients and mean free paths are related to each other via $\kappa = v \lambda /3$ where $v$ is the particle velocity, therefore the parallel diffusion coefficient within this strong scattering limit is given by $\kappa_{\parallel {\rm B}} = (3 p c v)/(q\:\delta B)$. Replacing the particle speed~$v$ by the particle momentum~$p$, this becomes
\begin{equation}
\kappa_{\parallel {\rm B}}(p) = \kappa_{\star} \: \frac{p^2}{(1+p^2)^{1/2}}
\label{eq:kB(p)}
\end{equation}
where momenta have been expressed in $m_p c$ units and $\kappa_{\star} = (3 m_p c^3)/(q\:\delta B)$. In our modeling the turbulence is not explicitly included and, therefore, the parameter $\delta B$ has to be seen as an external parameter. To stay close to previous work, in the following we assume that the turbulent field is approximately equal to the mean field $\delta B \approx B_0$. Then numerically we obtain
\begin{equation}
\kappa_{\star} = \frac{3\times 10^{22}}{B_0} \textnormal{cm}^2 \textnormal{s}^{-1}
\label{eq:kB_0}
\end{equation}
where $B_0$ is the value of the ambient field in microGauss. 

\subsection{A Simple Power-law Model}
\label{sec:models:powerlaw}

The Bohm limit described above is a simplified model for particle diffusion. \cite{2009APh....31..237S} and \cite{2014Ap&SS.350..197S} showed that it can be derived under the assumption of strong turbulence. In the weak turbulence limit, parallel diffusion is controlled by gyro-resonant interactions where the particles with a specific momentum only interact with a certain length scale of the turbulence (see, e.g., \citealt{2009ASSL..362.....S} for more details). In this case, the mean free path $\lambda_{\parallel}$ still increases with momentum, however it is no longer directly proportional to the unperturbed Larmor radius, but rather a power-law: $\lambda_{\parallel} \sim R_L^{\alpha}$ with $\alpha > 0$. The corresponding diffusion coefficient has the form $\kappa_{\parallel} \sim v R_L^{\alpha}$. We can recover the Bohm limit by setting $\alpha=1$. If particle diffusion can indeed be described by gyro-resonant interactions, $\alpha$ can have different values depending on the assumptions made for the turbulence spectrum. What the spectrum is at an SNR shock wave remains unclear. Furthermore, $\alpha$ can be different for different energies because the particles interact resonantly with different regimes of the turbulence \citep{2009ASSL..362.....S}. In the present paper we do not deal with details of turbulence theory and the different scattering processes. So in the following we simply consider the following form of the diffusion coefficient:
\begin{equation} 
\kappa_{\parallel}(p) = \kappa_{\star} \: p^{\alpha}
\label{eq:kPL(p)}
\end{equation}
where momentum is in $m_p c$ units and $\kappa_{\star}$ can be estimated from Equation~(\ref{eq:kB_0}). Note that in writing Equation~(\ref{eq:kPL(p)}) we considered relativistic particles for which $v=c$. As will be shown in the next section, we do not expect our model for perpendicular diffusion to be notably different at non-relativistic energies, and therefore for simplicity we do not consider these low energies in our study. We assume that particles are injected in the DSA mechanism and pre-accelerated up to $p \simeq m_p c$, a momentum that should be fairly easy to reach, and we investigate their subsequent fate. The simple, convenient form given by Equation~(\ref{eq:kPL(p)}) is commonly used in studies of DSA. It was in particular used in KRJ09, that we used as a testbed for our own explorations. 

We note that the Bohm coefficient given by Equation~(\ref{eq:kB(p)}) reduces to such a power-law as Equation~(\ref{eq:kPL(p)}) in both the non-relativistic and relativistic regimes: for $p \ll m_p c$, $\kappa_{\rm B}(p) \propto p^2$, whereas for $p \gg m_p c$, $\kappa_{\rm B}(p) \propto p^1$.

\subsection{A Realistic Model for Perpendicular Diffusion}
\label{sec:models:nonBohm}

\begin{figure}
\includegraphics[scale=1]{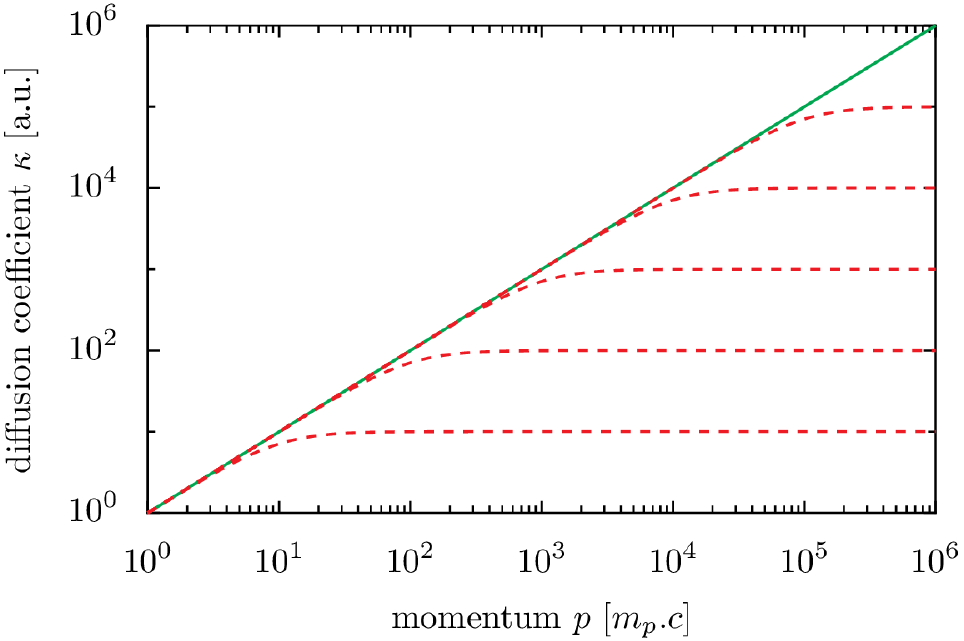}
\caption{The momentum dependence of the diffusion coefficients considered in this work. The solid green curve corresponds to the power-law case defined by Equation~(\ref{eq:kPL(p)}). The dashed red curves correspond to the modified cases defined by Equation~(\ref{eq:kNB(p)}), for different values of the characteristic momentum $p_{\rm c}$ (rising from $10\:m_p c$ to $10^5\:m_p c$ by factors of 10, from bottom to top). Here we have chosen the index $\alpha=1$.}
\label{fig:kappa}
\end{figure}

It has been historically recognized that the transport of particles across the magnetic field is an important, but poorly understood effect \citep[e.g.][]{1998SSRv...83..351G}. \cite{2010Ap&SS.325...99S} investigated the problem of perpendicular diffusion and found that the perpendicular mean free path increases with increasing particle momentum and that for high momenta the perpendicular mean free path becomes independent of the momentum. \cite{2014AdSpR..53.1024S} showed that this behavior of the perpendicular diffusion coefficient is universal (i.e., independent of the assumed turbulence properties). The momentum-dependence of the perpendicular diffusion coefficient was also explored numerically by using test-particle simulations. In \cite{2012AdSpR..49.1643Q}, for instance, the aforementioned behavior of the perpendicular diffusion coefficient was confirmed. Based on this previous work, we propose the following form for the perpendicular diffusion coefficient:
\begin{equation}
\kappa_{\perp}(p) = \kappa_{\infty} \: \frac{p^{\alpha}}{(p_{\rm c}^{2 \alpha}+p^{2 \alpha})^{1/2}}
\label{eq:kNB(p)}
\end{equation}
where momenta are again in $m_p c$ units, and with now three free parameters: the factor~$\kappa_{\infty}$, the index~$\alpha$, and a characteristic momentum $p_{\rm c}$. We see that for low momenta this coefficient behaves like a power-law, and that above the characteristic momentum it saturates to a constant: 
\begin{subnumcases}
{\kappa_{\perp}(p) = }
\kappa_{\infty} \left(\frac{p}{p_{\rm c}}\right)^{\alpha} & $p \ll p_{\rm c}$ \label{eq:kNB(p<<pc)} \\
\kappa_{\infty} & $p \gg p_{\rm c}$. \label{eq:kNB(p>>pc)}
\end{subnumcases}
Here again we have restricted the discussion to relativistic particles, and in particular we expect $p_{\rm c} > m_p c$. Analytical work (see \citealt{2010Ap&SS.325...99S} and \citealt{2013ApJ...774....7S}) has shown that $p_{\rm c}$ and $\kappa_{\infty}$ depend on turbulence parameters such as the magnetic field strength and the correlation length, but these parameters are basically unknown in SNRs. However, analytical theory predicts that for low particle momenta the perpendicular and parallel diffusion coefficient are related by $\kappa_{\perp} = 0.5 \left(\delta B / B_0\right)^2 \kappa_{\parallel}$, so that, under our assumption $\delta B \approx B_0$, we can match the alternative diffusion coefficient of Equation~(\ref{eq:kNB(p)}) with the common diffusion coefficient of Equation~(\ref{eq:kPL(p)}). That is, we set 
\begin{equation}
\kappa_{\infty} = \kappa_{\star} \: (p_{\rm c}^{2 \alpha}+p_{\rm inj}^{2 \alpha})^{1/2}, 
\label{eq:kNB_inf(p)}
\end{equation}
where $p_{\rm inj} < m_p c$ is the smallest momentum considered. Given this normalization, we note that our perpendicular diffusion coefficient will always be less than its Bohm counterpart. 
Having thus limited the number of free parameters, we will henceforth study the effect of varying~$p_{\rm c}$ for a given~$\alpha$. 
The resulting coefficients are plotted in Figure~\ref{fig:kappa} for comparison (for $\alpha=1$).

\subsection{Space Dependence}
\label{sec:models:space}

So far we have been concerned only with the momentum dependence of the diffusion coefficient, but the parameters defining~$\kappa$ can in principle vary in space, in particular the multiplicative pre-factor. Following KRJ09, we modify the form of the diffusion coefficient to account for the compression of magnetic field, assuming the mean field is tied to the density. This leads to
\begin{equation}
\kappa(x,p) = \left(\frac{\rho_0}{\rho(x)}\right)^{\nu} \kappa(p)
\label{eq:k(x)}
\end{equation}
where $\rho_0$ is the upstream (unperturbed) density, $\rho(x)$ is the local density at location~$x$, and $\nu$ is set to either~0 or~1 depending on whether the compression is taken into account or not.


\section{Analytical Considerations on the Maximum Momentum of Particles}
\label{sec:ana}

The maximum energy that can be reached by particles at SNR shocks is one of the major questions of the theory of CR production, and it ultimately comes down to how particles are interacting with magnetic fields. By confining particles in the vicinity of the shock front (where they can gain energy), the diffusion coefficient controls the timescale of the acceleration process. Starting from some injection momentum $p_{\rm inj}$,\footnote{The initial injection of particles in the DSA mechanism is a complex and still poorly understood process, that in principle depends on the obliquity of the shock. In this paper we simply assume that such a process is at work, and study the subsequent fate of particles.} the spectrum of accelerated particles extends progressively in time to higher and higher momenta, with a maximum momentum $p_{\rm max}$ that directly depends on the behaviour of~$\kappa(p)$. More precisely, from Equation~(3.31) in \citet{1983RPPh...46..973D}, the time it takes to accelerate particles from
$p_{\rm inj}$ to $p_{\rm max}$ is given by
\begin{equation}
\langle t\rangle_{\rm acc} = \frac{3}{u_1 - u_2} \int_{p_{\rm inj}}^{p_{\rm max}} \left(\frac{\kappa_1}{u_1} + \frac{\kappa_2}{u_2}\right) \frac{dp}{p}
\label{eq:tacc}
\end{equation}
with $u$ the flow velocity, $\kappa$ the diffusion coefficient, $p$ the particle momentum, and where the subscripts~1 and~2 indicate the upstream and downstream medium respectively (as defined in the shock frame). Using Equation~(\ref{eq:k(x)}) for the space dependence of~$\kappa$, we have $\kappa_2=\kappa_1$ if $\nu=0$ and $\kappa_2=\kappa_1/r$ if $\nu=1$, where we recall that~$r = \rho_2/\rho_1 = u_1/u_2$ is the compression ratio. Also, in the shock frame, we have $u_1=u_S$ and $u_2=u_S/r$ where $u_S$ is the shock speed. So Equation~(\ref{eq:tacc}) can be rewritten as
\begin{equation}
\langle t\rangle_{\rm acc} = \frac{R_{\nu}}{u_S^2} \int_{p_{\rm inj}}^{p_{\rm max}} \kappa_1(p)\: \frac{dp}{p}
\label{eq:tacc_uS}
\end{equation}
where $R_{\nu}$ is a dimensionless constant that can be expressed as a function of~$r$, in a way that depends on the behaviour assumed for $\kappa(x)$: 
\begin{subnumcases}
{R_{\nu}= }
\frac{3r(r+1)}{r-1} = 20 & $\nu=0$ \label{eq:R_nu=0} \\
\frac{6r}{r-1} = 8 & $\nu=1$ \label{eq:R_nu=1}
\end{subnumcases}
where for the numerical values we adopted $r=4$.

In the following, we compute the acceleration time for the different diffusion coefficients introduced in section~\ref{sec:models} (note that all momenta will be expressed in $m_p c$ units). When the integral has been computed, Equation~(\ref{eq:tacc_uS}) can be inverted to obtain an expression of $p_{\rm max}$ as a function of time~$t$ (for a given $p_{\rm inj}$). We note that these age-limited estimates are actually upper limits, since $p_{\rm max}$ can be limited by other processes. In particular, a general requirement is that the diffusion length of particles remains smaller than the size of the SNR -- otherwise particles are free to escape the accelerator without ever coming back to the shock. Another complication, for the case of perpendicular shocks, is that particles will drift along the shock surface \citep[e.g.][]{2013APh....43...56B}. Also, the maximum momentum of electrons will be limited by radiative losses since they emit synchrotron radiation efficiently. Our work in this paper is relevant to protons (or heavier ions).

\paragraph{Bohm diffusion coefficient.}
\label{sec:ana:Bohm}

Applying Equation~(\ref{eq:tacc_uS}) with Equation~(\ref{eq:kB(p)}), we obtain:
\begin{equation}
\langle t\rangle_{\rm acc \parallel B} = t_{0 \parallel} \left[
\sqrt{1+p_{\rm max}^2} - 
\sqrt{1+p_{\rm inj}^2}
\right]
\label{eq:tacc_B}
\end{equation}
where we have noted $t_{0 \parallel}=R_{\nu}\; \kappa_{\star} / u_S^2$.

\paragraph{Power-law diffusion coefficient.}
\label{sec:ana:powerlaw}

Applying Equation~(\ref{eq:tacc_uS}) with Equation~(\ref{eq:kPL(p)}), we obtain:
\begin{subnumcases}
{\langle t\rangle_{\rm acc \parallel} = t_{0 \parallel}}
\frac{p_{\rm max}^{\alpha} - p_{\rm inj}^{\alpha}}{\alpha} & $\alpha>0$ \label{eq:tacc_PL_a>0} \\
\ln\left(\frac{p_{\rm max}}{p_{\rm inj}}\right) & $\alpha=0$. \label{eq:tacc_PL_a=0}
\end{subnumcases}
Note that we expect the diffusion coefficient to be dependent on the particle momentum at least over some range, implying $\alpha>0$, however we recall that our new diffusion coefficient gets constant above the characteristic momentum~$p_{\rm c}$, so the case $\alpha = 0$ is of interest as well.
Equations~(\ref{eq:tacc_PL_a>0})--(\ref{eq:tacc_PL_a=0}) can be readily inverted as
\begin{subnumcases}
{\langle p_{\rm max}(t) \rangle_{\parallel} = }
p_{\rm inj} \left( 1 + \frac{\alpha}{p_{\rm inj}^\alpha } \frac{t}{t_{0 \parallel}} \right)^{1/\alpha} & $\alpha>0$ \label{eq:pmax_PL_a>0} \\
p_{\rm inj} \: \exp\left( \frac{t}{t_{0 \parallel}} \right) & $\alpha=0$ \label{eq:pmax_PL_a=0}.
\end{subnumcases}

\paragraph{Perpendicular diffusion coefficient.}
\label{sec:ana:nonBohm}

Applying Equation~(\ref{eq:tacc_uS}) with Equation~(\ref{eq:kNB(p)}), after using the substitution $x=p^\alpha$ to compute the integral, we obtain for $\alpha>0$:
\begin{eqnarray}
\langle t\rangle_{\rm acc \perp} & = & \frac{t_{0 \perp}}{\alpha} \left[
\sinh^{-1}\left( \left(\frac{p_{\rm max}}{p_{\rm c}}\right)^\alpha \right) - \sinh^{-1}\left( \left(\frac{p_{\rm inj}}{p_{\rm c}}\right)^\alpha \right)
\right]
\label{eq:tacc_NB_sinh} \\
& = & \frac{t_{0 \perp}}{\alpha} 
\ln\left( \frac{\sqrt{1+\left(\frac{p_{\rm max}}{p_{\rm c}}\right)^{2 \alpha}}+\left(\frac{p_{\rm max}}{p_{\rm c}}\right)^\alpha}{\sqrt{1+\left(\frac{p_{\rm inj}}{p_{\rm c}}\right)^{2 \alpha}}+\left(\frac{p_{\rm inj}}{p_{\rm c}}\right)^\alpha} \right)
\label{eq:tacc_NB_ln}
\end{eqnarray}
where we have noted $t_{0 \perp}=R_{\nu}\; \kappa_{\infty} / u_S^2$.
The case $\alpha = 0$ is similar to the power-law case, and is already covered by Equation~(\ref{eq:tacc_PL_a=0}) with $\kappa_{\star}$ replaced by $\kappa_{\infty}/\sqrt{2}$. The interesting case here is when $\alpha>0$. When $p_{\rm c} \ll p_{\rm inj} < p_{\rm max}$ 
Equation~(\ref{eq:tacc_NB_ln}) also reduces to the form of Equation~(\ref{eq:tacc_PL_a=0}), in agreement with the fact that the diffusion coefficient is then constant at all momenta. In the other limit $p_{\rm c} \gg p_{\rm max} > p_{\rm inj}$ 
Equation~(\ref{eq:tacc_NB_ln}) reduces to the form of Equation~(\ref{eq:tacc_PL_a>0}), in agreement with the fact that the diffusion coefficient is then a power-law of index~$\alpha$ at all momenta. In the intermediate case $p_{\rm inj} \le p_{\rm c} \le p_{\rm max}$, we can make the simplifying assumption that $\kappa_{\perp}(p)$ is a power-law for all~$p$ up to $p_{\rm c}$ and then a constant for all~$p$ above~$p_{\rm c}$ (see Equations~(\ref{eq:kNB(p<<pc)})--(\ref{eq:kNB(p>>pc)})). Then, adapting Equations~(\ref{eq:tacc_PL_a>0}) and~(\ref{eq:tacc_PL_a=0}) we obtain
\begin{subnumcases}
{\langle t\rangle_{\rm acc \perp} \simeq }
\frac{t_{0 \perp}}{\alpha} \left[ \left(\frac{p_{\rm max}(t)}{p_{\rm c}}\right)^{\alpha} - \left(\frac{p_{\rm inj}}{p_{\rm c}}\right)^{\alpha} \right] & 
$\begin{array}{l} p_{\rm max}(t) \le p_{\rm c} \\ \\ \\ \end{array}$ \label{eq:tacc_NB(p_max<pc)} \\
t_{\rm c} + t_{0 \perp}\: \ln\left(\frac{p_{\rm max}(t)}{p_{\rm c}} \right) & 
$\begin{array}{l} p_{\rm max}(t) \ge p_{\rm c} \\ \\ \\ \end{array}$ \label{eq:tacc_NB(p_max>pc)}
\end{subnumcases}
where we have noted $t_{0 \perp} = {R_{\nu}\; \kappa_{\infty}}/{u_S^2}$ and 
\begin{equation}
t_{\rm c} = \frac{t_{0 \perp}}{\alpha} \left(1 - \left(\frac{p_{\rm inj}}{p_{\rm c}}\right)^\alpha \right)
\label{eq:tc}
\end{equation}
the acceleration time from $p_{\rm inj}$ to $p_{\rm c}$. Note that, given that $p_{\rm inj} \ll p_{\rm c}$, $t_{\rm c}$~happens to be very weakly dependent on $p_{\rm c}$: $t_{\rm c} \simeq t_{0 \perp}/\alpha$.
We can invert Equations~(\ref{eq:tacc_NB(p_max<pc)}) and~(\ref{eq:tacc_NB(p_max>pc)}) (or adapt Equations~(\ref{eq:pmax_PL_a>0}) and~(\ref{eq:pmax_PL_a=0})) to obtain an estimate of the maximum momentum at time~$t$:
\begin{subnumcases}
{\langle p_{\rm max}(t) \rangle_{\perp} \simeq  }
p_{\rm inj}  \left( 1 + \alpha \left(\frac{p_{\rm c}}{p_{\rm inj}}\right)^\alpha \frac{t}{t_{0 \perp}} \right)^{1/\alpha} & 
$\begin{array}{l} t \le t_{\rm c}\\ \\ \\ \end{array}$ \label{eq:pmax_NB(t<tc)} \\
p_{\rm c} \exp\left( \frac{t-t_{\rm c}}{t_{0 \perp}} \right) & 
$\begin{array}{l} t \ge t_{\rm c}\\ \\ \\ \end{array}$. \label{eq:pmax_NB(t>tc)}
\end{subnumcases}

\begin{figure}
\includegraphics[scale=1]{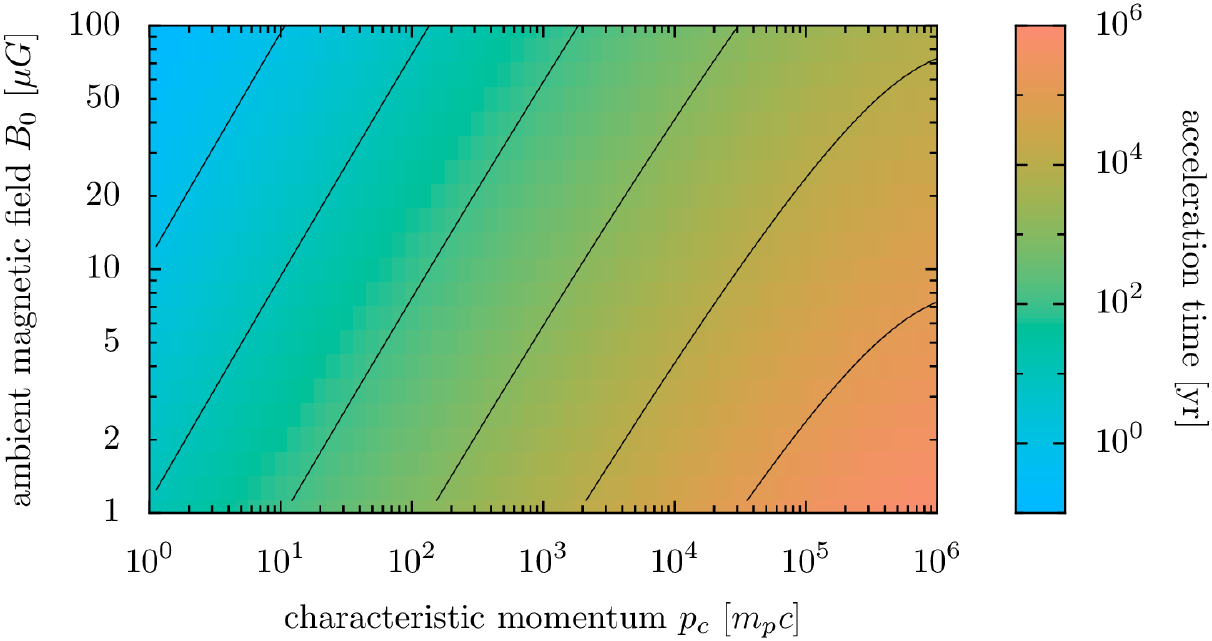}
\caption{Acceleration time to the ``knee" ($p_{\rm max}=10^6\:m_p c$) from the thermal pool ($p_{\rm inj}=10^{-2}\:m_p c$) for a shock of constant speed $u_S=1000$ km/s. We used the diffusion coefficient~(\ref{eq:kNB(p)}), which is a function of the characteristic momentum $p_{\rm c}$, and of the ambient magnetic field $B_0$ through Equations~(\ref{eq:kNB_inf(p)}) and~(\ref{eq:kB_0}). The acceleration time is computed from Equation~(\ref{eq:tacc_NB_sinh}) and is represented in colours, ranging from less than a year in blue to one million years in red (iso-contours have been added for each power of ten).}
\label{fig:tacc}
\end{figure}

The preceding formulae allow us to make order of magnitude estimates for SNRs under different scenarios. They will be useful to set the simulation time in the next section. In Figure~\ref{fig:tacc} we plot the acceleration time from $p_{\rm inj}=10^{-2}$ $m_p c$ to $p_{\rm max} = 10^6$  $m_p c$, assuming a shock of constant velocity $u_S=1000$ km/s (in reality $u_S$ would continuously decrease over time). We have used the diffusion coefficient~(\ref{eq:kNB(p)}) and set $\alpha=1$. The standard case of section~\ref{sec:models:powerlaw} is recovered when $p_{\rm c}$ approaches $p_{\rm max} = 10^6$ $m_p c$. We see that, in this case, the ``knee" energy can be reached in reasonable times (of the order of a thousand years) only if the magnetic field at the shock reaches values far above the typical Galactic value of a few $\mu G$. But if the value of $p_{\rm c}$ is lower, so that the diffusion coefficient becomes constant at high momenta, then acceleration proceeds much faster, and it gets much easier to reach the knee -- regardless of the value of $B_0$. 
We note that it had already been pointed out by \cite{1987ApJ...313..842J} that, if the perpendicular diffusion coefficient were much smaller than the parallel one but non negligible, the maximum energy attainable would be much higher than usually assumed (see also \citealt{2006A&A...454..687M}). Our proposed diffusion coefficient achieves this. 
However, we recall that the acceleration time scales directly with the unknown parameter~$\kappa_{\infty}$ (see Equations~(\ref{eq:tacc_NB_sinh})--(\ref{eq:tacc_NB_ln})).

We note that the acceleration times and the maximum momenta computed in this section were denoted in brackets, meaning that they represent the most probable values. The DSA mechanism is stochastic in nature, and although the advection in momentum always proceeds upwards, not all particles  injected initially will reach the same momentum at the same time. As a result, the spectrum will not show a sharp cut-off at $p_{\rm max}$, but rather an exponential cut-off \citep{1983RPPh...46..973D}. The actual shape of the cut-off depends on the diffusion coefficient: considering a power-law of index~$\alpha$, then the higher~$\alpha$, the more distinct the diffusive behaviours of particles of different momenta is, and the sharper the cut-off is. On the contrary, the new model that we propose has a plateau of the diffusion properties at high momenta, and so we expect that the cut-off around $p_{\rm max}$ will be less clearly defined than in standard studies of DSA. 
Finally, note that, as long as DSA operates in the linear regime -- that is, as long as the presence of accelerated particles doesn't modify the shock that accelerates them, the shape of the spectrum \emph{between} the injection domain ($p \gtrsim p_{\rm inj}$) and the cut-off domain ($p \lesssim p_{\rm max}$) does \emph{not} depend on $\kappa(p)$: it is always a power-law in momentum, $f(p) \propto p^s$, where the index $s$ depends on the compression ratio of the shock only (according to $s=3r/(r-1)$). But when one takes into account the back-reaction of particles on the shock (non-linear DSA), then the spectrum becomes more complicated, and can bear the mark of the diffusion process \citep{1983RPPh...46..973D,1991SSRv...58..259J,2001RPPh...64..429M}. To model the complex interplay between energetic particles and the shock, we then have to resort to numerical simulations \citep{2001JKAS...34..231J}.


\section{Numerical Simulations of the Spectrum of Accelerated Particles}
\label{sec:num}

When particles are efficiently accelerated, their pressure can become high enough that it affects the evolution of the shock, which in turn affects the way they are accelerated. As energetic particles diffuse upstream of the shock front, they progressively slow down the incoming flow (in the shock frame), which results in the formation of a characteristic \emph{precursor} in the upstream medium. As particles of different momenta travel over different distances away from the shock discontinuity (now referred to as the \emph{sub-shock}), they will experience different compression ratios, leading to a varying spectral index, which can no longer be described by a pure power-law -- a key prediction of non-linear DSA \citep{1991SSRv...58..259J,2001RPPh...64..429M}. On top of this, there exists another back-reaction loop, exerted through the magnetic field: energetic particles are thought to trigger instabilities while they stream in the upstream medium, thereby generating themselves the magnetic turbulence they need to be confined in the vicinity of the shock \citep{2013APh....43...56B}. In this work we do not address the details of magnetic field amplification, which are quite complex and fairly uncertain (see \citealt{2012SSRv..173..491S} for a review). An MHD treatment is beyond the scope of this paper. We simply assume the presence of magnetic waves that can scatter the particles, so that the DSA approach is valid. The main effect of increasing the magnetic field strength would be to alter the parameters of the diffusion coefficient (pre-factor $\kappa_{\rm \star}$ or $\kappa_{\rm \infty}$, characteristic momentum $p_{\rm c}$), which are essentially unconstrained anyway, and that we consider as free parameters. Our focus here is on the \emph{momentum-dependence} of~$\kappa$, and how it impacts the spectrum of particles in comparison with standard cases. From the previous discussion, $\kappa(p)$ clearly has a major role to play in the non-linear regime. Upstream of the shock the balance between advection and diffusion leads to an exponential decay of the particle density \citep[e.g.][]{1983RPPh...46..973D}, with a length scale
\begin{equation}
x_{\rm diff}(p) = \frac{\kappa(p)}{u_S}
\label{eq:x_diff}
\end{equation}
where $u_S$ is the shock velocity. Expression~(\ref{eq:x_diff}) can be interpreted as the typical extent of the region that particles of momentum~$p$ diffusively explore upstream of the shock.
For the standard (relativistic) Bohm diffusion model, where $\kappa(p)$ and thus $x_{\rm diff}(p)$ are directly proportional to~$p$, the spectrum gets distinctively concave. For our proposed perpendicular diffusion model, where $\kappa(p)$ and thus $x_{\rm diff}(p)$ eventually become independent of~$p$, we expect a different outcome. 

In this section, we investigate this effect by means of numerical simulations. Several models have been developed to simulate DSA, depending on the focus of the study. Here we are interested in the shape of the CR spectrum, so we are using a kinetic description for the particles: we work with the distribution function $f(x,p,t)$, which obeys the transport Equation~(\ref{eq:transport}), that we are numerically integrating in time in a way similar to KRJ09 (see \citealt{2010MNRAS.407.1773C} for a comparison of different possible kinetic approaches to solve this equation). On the other hand, for now we want to abstract the results from any environmental effects that may affect the shock evolution, and so we will consider an idealized shock front. 

\subsection{The code {\sc Marcos}}
\label{sec:num:marcos}

To perform such hydro-kinetic simulations of the coupled shock+particles system, we employ the existing code {\sc Marcos} \citep{2008MNRAS.383...41F}, developed for the study of time-dependent non-linear DSA. We recall here briefly the code's main features and refer the reader to the original paper \citep{2008MNRAS.383...41F} for more details. 

The code jointly solves in time the three conservation equations that describe the thermal fluid (Euler equations) and the advection-diffusion equation~(\ref{eq:transport}) that describes the energetic particles. The hydro equations, written in conservative form for ($\rho$, $\rho u$, $E$), are solved using an explicit Godunov scheme. The kinetic equation, written for $g=p^{4}\, f$ as a function of $y=\ln p$, is solved using a semi-implicit Crank-Nicolson scheme, in order to deal with the much smaller time scales associated with diffusion than with advection. In the version used for this paper, the code includes the effect of Alfv{\'e}nic heating on the flow and of Alfv{\'e}nic drift on the particles \citep[e.g.][]{2013JKAS...46...49K}. The geometry is 1D plane-parallel in space~$x$ (so the shock is unidirectional) and 1D spherically symmetric in momentum~$p$ (so the particle distribution is isotropic). The magnetic field is not explicitly computed by the code, it only appears as a parameter embedded in the diffusion coefficient. So the inclination of the shock normal with respect to the local magnetic field is defined implicitly by the choice of the diffusion model: when using Equation~(\ref{eq:kB(p)}) or~(\ref{eq:kPL(p)}) we are dealing with a purely parallel shock, when using Equation~(\ref{eq:kNB(p)}) we are dealing with a purely perpendicular shock. We note that the actual shock inclination in SNRs is much uncertain, and probably varies in a given object; we are here considering these two idealized cases separately. In order to accommodate the potentially huge range of scales caused by diffusion, we use AMR: the space grid is adaptively refined by deploying a hierarchy of nested grids of increasing resolution around the shock. To further reduce the cost of computing the acceleration of particles to ever-increasing momenta, we use parallelization: the momentum grid can be split over several processors. 

The shock is generated by a constant velocity piston. In practice, we are working in the piston frame to save on grid space, and so the piston appears as a fixed, reflecting wall at the left boundary of the grid. The position and properties of the shock front are diagnosed at each hydro time step. There, particles are injected continuously in time from the thermal pool, for simplicity we consider that this occurs in a fixed momentum bin $p_{\rm inj}$ and at a constant rate~$\eta$ (expressed as a fraction of the flux crossing the shock). The DSA solver then spreads the distribution of particles over space and momentum. The pressure of particles is computed from their momentum distribution in each space cell, and is included in the hydro equations  at each time step, leading to modified shocks and spectra. 

\subsection{Parameters}
\label{sec:num:parameters}

We aim to simulate DSA with parameters relevant for SNRs. We based our models on the work by KRJ09 that we used to validate and benchmark our own code. As KRJ09 we consider a shock of Mach number $M_{\rm S}=10$. Although the Mach number can be much higher in young SNRs, this is high enough to get a strong shock with compression of almost~4, and this is more computationally tractable. We concentrate on KRJ09 ``T6" models, which are relevant for a hot and tenuous ISM phase (such as inside a stellar wind bubble or a superbubble), since they were easier to reproduce. That is, we use an upstream temperature $T_0=10^6$~K and an upstream particle density $n_0=0.03$~cm$^{-3}$. For $M_{\rm S}=10$ this corresponds to a reasonable shock velocity $u_{\rm S}\simeq1500$~km.s$^{-1}$ (assuming a mean molecular weight $\mu=0.6$). We consider an ambient magnetic field $B_0=5~\mu$G, appropriate for the average Galactic field, which according to Equation~(\ref{eq:kB_0}) leads to $\kappa_{\star} = 6.23\times 10^{21}$ cm$^2$s$^{-1}$, as in KRJ09. Regarding the momentum dependence of the diffusion coefficient, we compare a standard power-law of index~$\alpha$, Equation~(\ref{eq:kPL(p)}), with our modified version of same index~$\alpha$ and varying characteristic momentum $p_{\rm c}$, Equation~(\ref{eq:kNB(p)}). For $\alpha$ we consider the two limit cases used by KRJ: $\alpha=1$, which is the relativistic Bohm scaling, and $\alpha=1/2$. For each~$\alpha$, we vary the parameter $p_{\rm c}$ between $m_pc$ and the maximum momentum considered. We model momenta within the range from $10^{-2}$ to $10^4$ $m_pc$ for $\alpha=1$ and from $10^{-2}$ to $10^8$ $m_pc$ for $\alpha=1/2$. We use an injection momentum $p_{\rm inj} = 10^{-2}\:m_p c$, and an injection fraction $\eta = 10^{-4}$ or $\eta = 10^{-3}$, consistent with KRJ09. Regarding the space dependence of the diffusion coefficient, we opt to tie it to the density for the case $\alpha=1$ ($\nu=1$, models suffixed ``d" in KRJ09), and to take it uniform for the case $\alpha=1/2$ ($\nu=0$). The kinetic parameters for the two sets of models are summarized in Table~\ref{tab:params}.

\begin{table*}[ht]
\begin{centering}
\begin{tabular}{|c|c|c|c|c|c|c|}
\hline 
$\alpha$ & $\kappa_{\star}$ & $\nu$ & $\eta$ & $p_{\rm inj}$ & $p_{\rm c}$ & $p_{\rm max}$\tabularnewline
\hline 
\hline 
1 & $6.23\times10^{21}$ cm$^{2}$.s$^{-1}$ & 1 & $10^{-4}$ or $10^{-3}$ & $10^{-2}\, m_{p}c$  & $10^{0},\,10^{1},\,10^{2},\,10^{3}\, m_{p}c$  & $10^{4}\, m_{p}c$ \tabularnewline
\hline 
1/2 & $6.23\times10^{21}$ cm$^{2}$.s$^{-1}$ & 0 & $10^{-4}$ or $10^{-3}$ & $10^{-2}\, m_{p}c$  & $10^{0},\,10^{2},\,10^{4},\,10^{6}\, m_{p}c$  & $10^{8}\, m_{p}c$ \tabularnewline
\hline 
\end{tabular}
\par\end{centering}
\caption{Kinetic parameters for the two sets of models. Particles are injected at the shock front at a constant rate~$\eta$ at a fixed momentum $p_{\rm inj}$. They experience DSA with a diffusion coefficient determined by parameters~$\alpha$, $p_{\rm c}$, $\kappa_{\star}$, and~$\nu$ according to equations~(\ref{eq:kNB(p)}), (\ref{eq:kNB_inf(p)}), and~(\ref{eq:k(x)}). The simulation is run until they reach the target momentum $p_{\rm max}$.
\label{tab:params}}
\end{table*}

\begin{table*}[ht]
\begin{centering}
\begin{tabular}{|c|c|c|c|}
\hline 
$p_{\mathrm{c}}$ for $\alpha=1/2$ & $p_{\mathrm{c}}$ for $\alpha=1$ & length unit & time unit\tabularnewline
\hline 
\hline 
$>10^{8}\, m_{p}c$ & $>10^{4}\, m_{p}c$ & $4.1\times10^{17}$~cm= $1.4\times10^{-1}$~pc & $2.7\times10^{9}$~s = $8.6\times10^{+1}$~yr\tabularnewline
\hline 
$10^{6}\, m_{p}c$ & $10^{3}\, m_{p}c$ & $4.1\times10^{16}$~cm= $1.4\times10^{-2}$~pc & $2.7\times10^{8}$~s = $8.6\times10^{-0}$~yr\tabularnewline
\hline 
$10^{4}\, m_{p}c$ & $10^{2}\, m_{p}c$ & $4.1\times10^{15}$~cm= $1.4\times10^{-3}$~pc & $2.7\times10^{7}$~s = $8.6\times10^{-1}$~yr\tabularnewline
\hline 
$10^{2}\, m_{p}c$ & $10^{1}\, m_{p}c$ & $4.1\times10^{14}$~cm= $1.4\times10^{-4}$~pc & $2.7\times10^{6}$~s = $8.6\times10^{-2}$~yr\tabularnewline
\hline 
$10^{0}\, m_{p}c$ & $10^{0}\, m_{p}c$ & $4.1\times10^{13}$~cm= $1.4\times10^{-5}$~pc & $2.7\times10^{5}$~s = $8.6\times10^{-3}$~yr\tabularnewline
\hline 
\end{tabular}
\par\end{centering}
\caption{Code units of length and time for the different values of~$p_{\mathrm{c}}$, as defined by equation~(\ref{eq:kNB(p)}), increasing from bottom to top (the top row corresponds to the reference case of equation~(\ref{eq:kPL(p)})). Note that values have been grouped by equivalent pairs for the cases $\alpha=1,1/2$.
\label{tab:code_units}}
\end{table*}

We use code units as in KRJ09. In these units the upstream mass density is normalized to unity, the shock velocity is normalized to unity, and the diffusion coefficient is normalized to its value at the maximum momentum $p_{\rm max}$ (with the choice of normalization Equation~(\ref{eq:kNB_inf(p)}), this implies that for our new diffusion law $\kappa_{\infty}$ is always of order unity for $p_{\rm c} \ll p_{\rm max}$, whereas for a pure power-law $\kappa_{\star}$ scales as $p_{\rm max}^{-\alpha}$). These conditions implicitly define the units of mass, length and time (the latter two are reported for convenience in Table~\ref{tab:code_units}). Note that independently of this, particle momenta are always expressed in units of $m_p c$. We can estimate the time over which to run our simulations according to the relevant equations in section~\ref{sec:ana}. For the standard power-law, Equation~(\ref{eq:tacc_PL_a>0}) reduces for $p_{\rm max} \gg p_{\rm inj}$ to $\langle t\rangle_{\rm acc \parallel} = R_{\nu} / \alpha \: \kappa(p_{\rm max}) / u_S^2$, which in our code units simply reads $R_{\nu} / \alpha$, where we recall that $R_{\nu}$ is given by Equations~(\ref{eq:R_nu=0})--(\ref{eq:R_nu=1}). For our new diffusion law, Equation~(\ref{eq:tacc_NB(p_max>pc)}) reduces for $p_{\rm c} \gg p_{\rm inj}$ to $\langle t\rangle_{\rm acc \parallel} = R_{\nu} \: \kappa(p_{\rm max}) / u_S^2 \:(1/\alpha + \ln(p_{\rm max}/p_{\rm c}))$, which in our code units reads $R_{\nu} \:(1/\alpha + \ln(p_{\rm max}/p_{\rm c}))$ (note that we recover the previous case by setting $p_{\rm max}=p_{\rm c}$). It is important to note that, even though the simulation time in code units gets longer with the perpendicular diffusion coefficient, the simulation time in physical units gets shorter (see Table~\ref{tab:code_units}), because with our choice of normalization we always have $\kappa_{\perp} \le \kappa_{\parallel}$.
The size of the space domain is set so that it contains the shock over the time simulated, and also the precursor formed by the particles. The scale of the precursor is given by the diffusion length upstream of the shock of particles of the highest momentum: $x_{\rm diff}(p_{\rm max})$ (see Equation~(\ref{eq:x_diff})), which in our code units is conveniently of unity. A~final important point to be discussed regarding the simulations design is the space resolution~${\rm d}x$. In order to resolve the diffusive behaviour of the particles, each space cell must be smaller than the diffusion length of all particles present in this region. At the shock front this means that we should have ${\rm d}x \ll x_{\rm diff}(p_{\rm inj})$. In practice we use ${\rm d}x / x_{\rm diff}(p_{\rm inj}) = 1/3$ for all simulations (for $\alpha=1/2$ this is almost the same resolution as KRJ09's ``T6P12" model, for $\alpha=1$ this is a much higher resolution than KRJ09's ``T6P1d'' model, which is needed for our code to give converged results).

\subsection{Results for $\alpha=1$}
\label{sec:num:alpha1}

\begin{figure}
\includegraphics[scale=0.7]{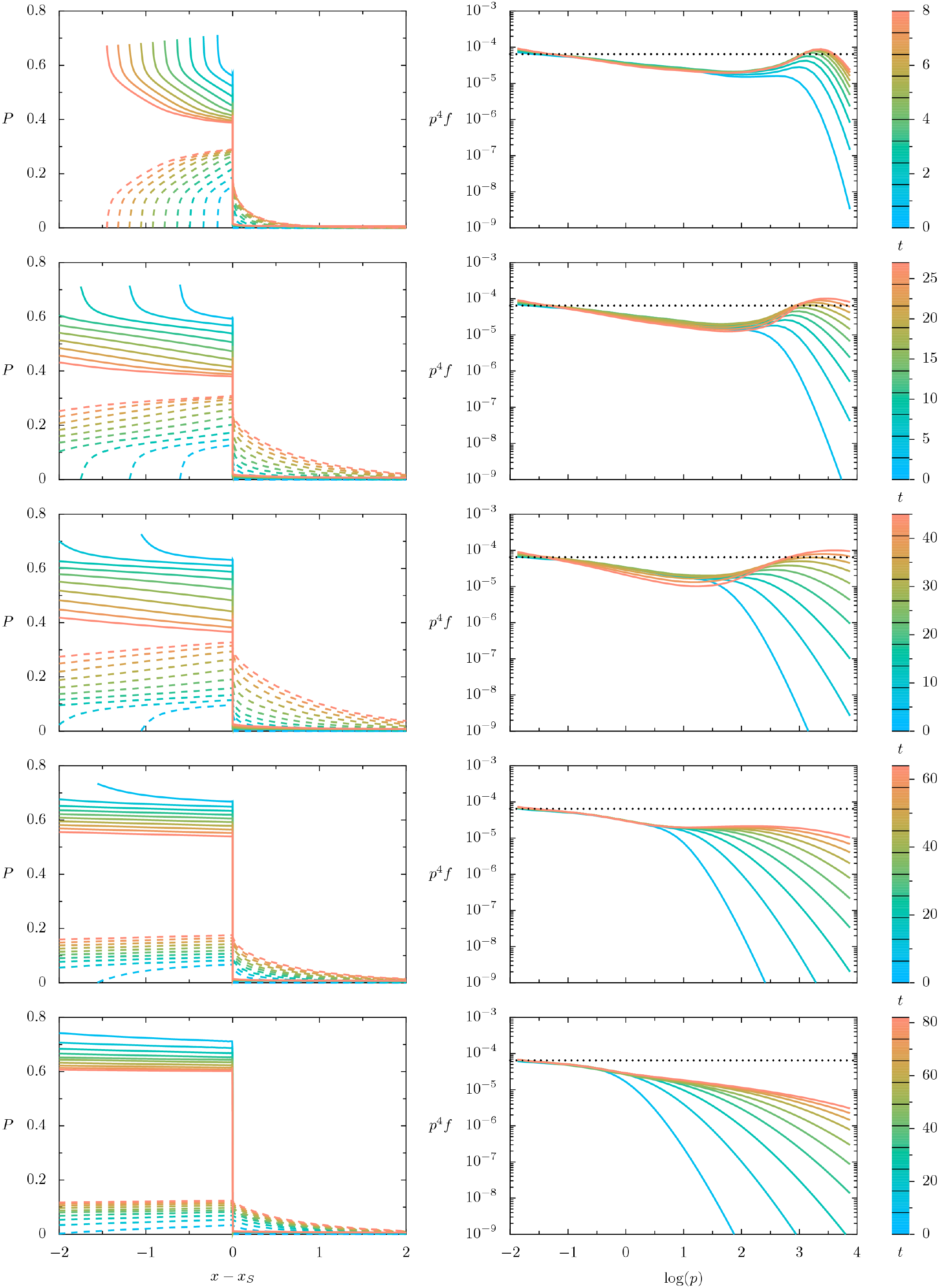}
\caption{Left: Pressure profiles (solid: fluid, dashed: particles) as a function of position at different times.  Right: spectrum of the particles ($g=p^4 f$) as a function of momentum at different times. Diffusion with $\alpha=1$ and $p_{\rm c}$ increasing from bottom to top: $10^0$ $m_pc$, $10^2$ $m_pc$, $10^4$ $m_pc$, $10^6$ $m_pc$, and $>10^8$ $m_pc$ (the reference Bohm case). Injection level $\eta=10^{-4}$.}
\label{fig:T6P1_hydro_kino_1e-4}
\end{figure}

\begin{figure}
\includegraphics[scale=0.7]{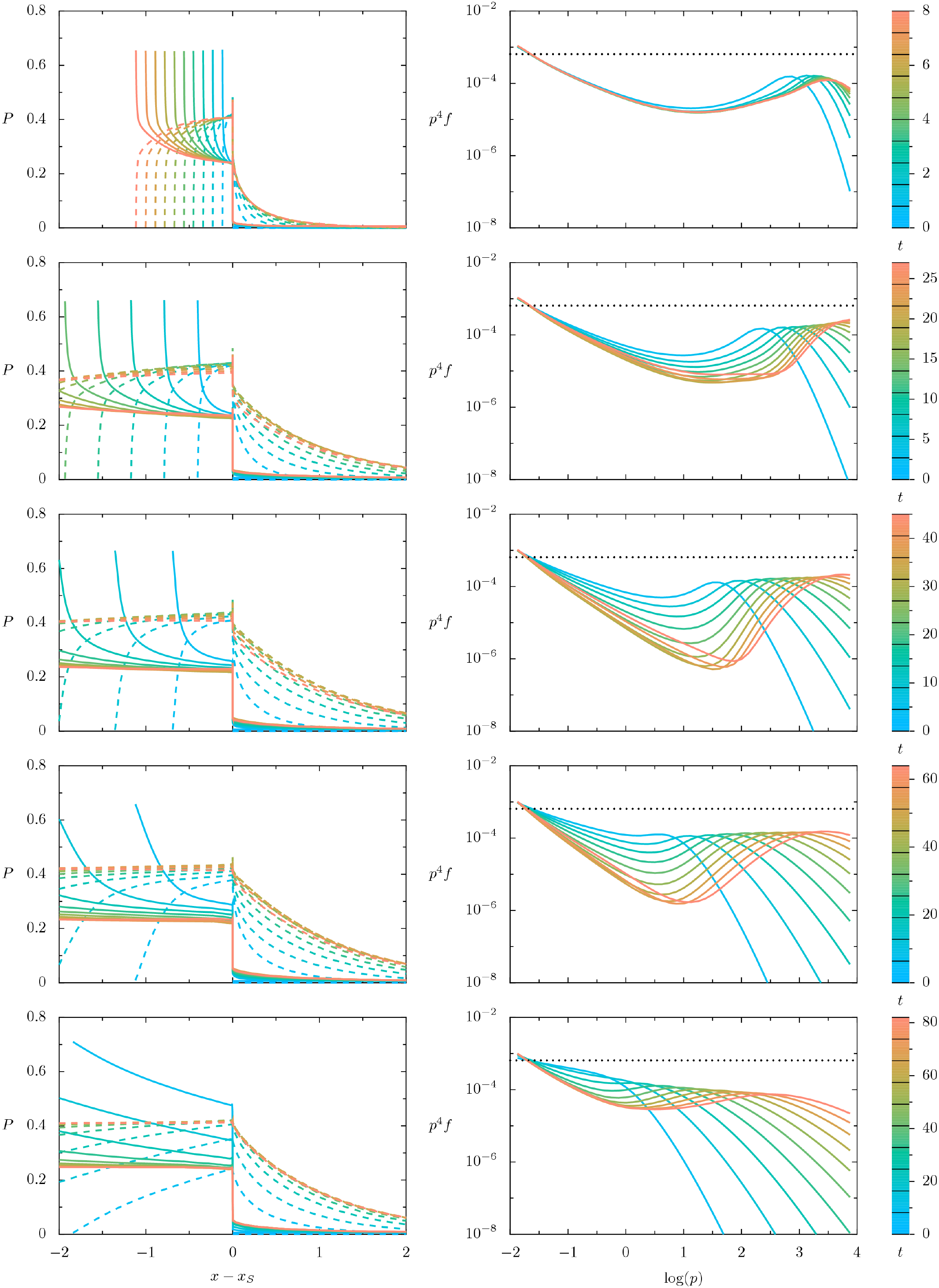}
\caption{Left: Pressure profiles (solid: fluid, dashed: particles) as a function of position at different times.  Right: spectrum of the particles ($g=p^4 f$) as a function of momentum at different times. Diffusion with $\alpha=1$ and $p_{\rm c}$ increasing from bottom to top: $10^0$ $m_pc$, $10^2$ $m_pc$, $10^4$ $m_pc$, $10^6$ $m_pc$, and $>10^8$ $m_pc$ (the reference Bohm case). Injection level $\eta=10^{-3}$.}
\label{fig:T6P1_hydro_kino_1e-3}
\end{figure}

We start our study of the properties of our perpendicular diffusion model with the common choice $\alpha=1$. The simulation results are presented in Figure~\ref{fig:T6P1_hydro_kino_1e-4} for a low injection fraction $\eta=10^{-4}$ and Figure~\ref{fig:T6P1_hydro_kino_1e-3} for a high injection fraction $\eta=10^{-3}$. On each figure, the left column shows a hydrodynamic quantity: the pressure profiles~$P$ as a function of position~$x$ along the direction of the shock propagation, and the right column shows a kinetic quantity: the particle spectrum $g = p^4 f$ as a function of momentum~$p$. The results are plotted for various values of the characteristic momentum, increasing from bottom to top: $p_{\rm c}$ = $10^0$, $10^1$, $10^2$, $10^3$ $m_p c$, and the topmost plots correspond to the reference Bohm case, which formally is recovered with our new coefficient when $p_{\rm c} > p_{\rm max} = 10^4\:m_p c$. Finally, on each plot, the quantity of interest is plotted at different times in different colors, increasing from the startup time in blue to the final time in red (according to the color bar on the right of each row). 

Pressure profiles (on the left columns) are plotted in the shock frame, since we are not interested in the extent traveled by the shock, but rather by the shape of the shock front. The shock is visible as a sharp discontinuity in all profiles, with a compression ratio~$r$ of about~4 for the density (not shown) as expected. As time passes, we clearly see the formation of a precursor upstream of the shock front. As already explained, this is caused by the presence of energetic particles that are free to wander ahead of the shock. The pressure of the particles has been added as dashed curves. One can see that the particle pressure is building up in the upstream region, and that in the downstream region it quickly rises to a level comparable with the fluid pressure, while the fluid pressure decreases accordingly (note that in our units the pressure is normalized by the upstream ram pressure). The intensity of this back-reaction effect depends on the injection level: for $\eta=10^{-4}$ (Figure~\ref{fig:T6P1_hydro_kino_1e-4}) the particle pressure always remains lower than the fluid pressure, whereas for $\eta=10^{-3}$ (Figure~\ref{fig:T6P1_hydro_kino_1e-3}) the particle pressure dominates. When it comes to the diffusion properties, we see that the results are qualitatively similar for the different values of~$p_{\rm c}$. The final share between the fluid and particle pressures varies very slightly for $\eta=10^{-3}$, more appreciably for $\eta=10^{-4}$, but the overall dynamics are the same.

Particle spectra (on the right columns) are plotted immediately downstream of the shock front. As time passes, we see how the spectrum extends to higher and higher momenta (in a way that broadly agrees with the analytical estimates of section~\ref{sec:ana}). In this $\log(g)$--$\log(p)$ representation, it would stay perfectly flat in the test-particle regime (dotted line). Because of the shock modifications discussed above, it actually comes in a variety of shapes. In the case where the diffusion coefficient is a strictly increasing function of momentum (topmost plots), particles of low momenta remain in the vicinity of the shock where they experience a compression of less than~4, leading to a slope $s>4$, whereas particles of high momenta explore the whole shock structure and experience a compression of more than~$4$, leading to a slope $s<4$. This results in a characteristic curved spectrum for the Bohm diffusion. When turning to our new diffusion coefficient, a more complex behaviour is observed as a function of the characteristic momentum~$p_{\rm c}$. For a value of~$p_{\rm c}$ close to $p_{\rm max}$, such as $p_{\rm c} = 10^6$ $m_p c$ (second row from the top), the results are similar to the Bohm case, because as long as $p < p_{\rm c}$ our new~$\kappa$ reduces to the same power-law of index~$\alpha$. However, as $p_{\rm c}$ decreases (from top to bottom rows), the spectrum changes at high momenta, because particles reaching these momenta experience a different diffusive behaviour (the regime where $\kappa$ becomes constant). The upward curvature before $p_{\rm max}$ becomes less and less pronounced. For a low injection fraction $\eta=10^{-4}$, it eventually completely disappears. For a high injection fraction $\eta=10^{-3}$, it remains, but much reduced and shifted towards lower momenta. This behavior can be understood as follows: when the diffusion coefficient becomes independent of the particle momentum, the diffusion length of particles upstream of the shock (Equation~(\ref{eq:x_diff})) becomes constant, and therefore they explore essentially the same region in space and experience the same compression, leading to the same spectral index, and therefore to a rather straight spectrum above $p_{\rm c}$. 
By~comparing the spectrum at different times (in particular for the cases with a low $p_{\rm c}$, such as $p_{\rm c}=1$~$m_p c$ on the bottom row), we also see how the diffusion coefficient affects the shape of the cut-off around $p_{\rm max}$. At early times, when $p_{\rm max}$ is still lower than $p_{\rm c}$, the cut-off is quite sharp, whereas at later times, when $p_{\rm max}$ is higher than $p_{\rm c}$, the cut-off broadens. Also note that the output times were chosen regularly spaced along the course of the simulation. As a result, the spacing of curves highlights the acceleration timescales. In the standard diffusion case, it is apparent that most of the time is taken to accelerate particles over the very last decade of the momentum box. In the modified case, the same behaviour is observed until~$p_{\rm max}$ reaches~$p_{\rm c}$, and then the acceleration becomes more uniform over time (in the constant diffusion regime, $\log(p_{\rm max})$ grows linearly with time, see Equation~(\ref{eq:pmax_NB(t>tc)})).

\subsection{Results for $\alpha=1/2$}
\label{sec:num:alpha12}

\begin{figure}
\includegraphics[scale=0.7]{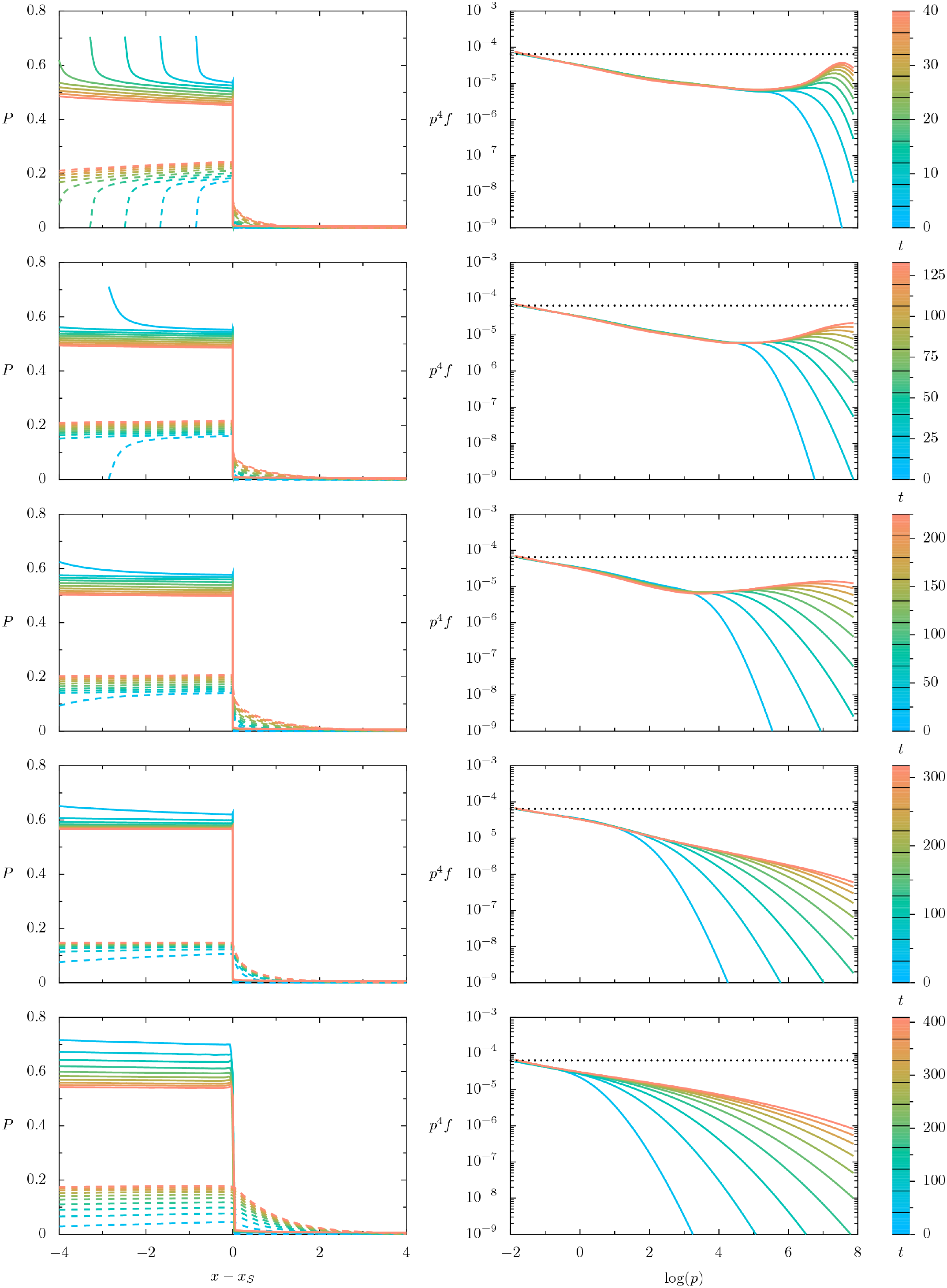}
\caption{Left: Pressure profiles (solid: fluid, dashed: particles) as a function of position at different times.  Right: spectrum of the particles ($g=p^4 f$) as a function of momentum at different times. Diffusion with $\alpha=1/2$ and $p_{\rm c}$ increasing from bottom to top: $10^0$ $m_pc$, $10^2$ $m_pc$, $10^4$ $m_pc$, $10^6$ $m_pc$, and $>10^8$ $m_pc$ (the reference Bohm case). Injection level $\eta=10^{-4}$.}
\label{fig:T6P12_hydro_kino_1e-4}
\end{figure}

\begin{figure}
\includegraphics[scale=0.7]{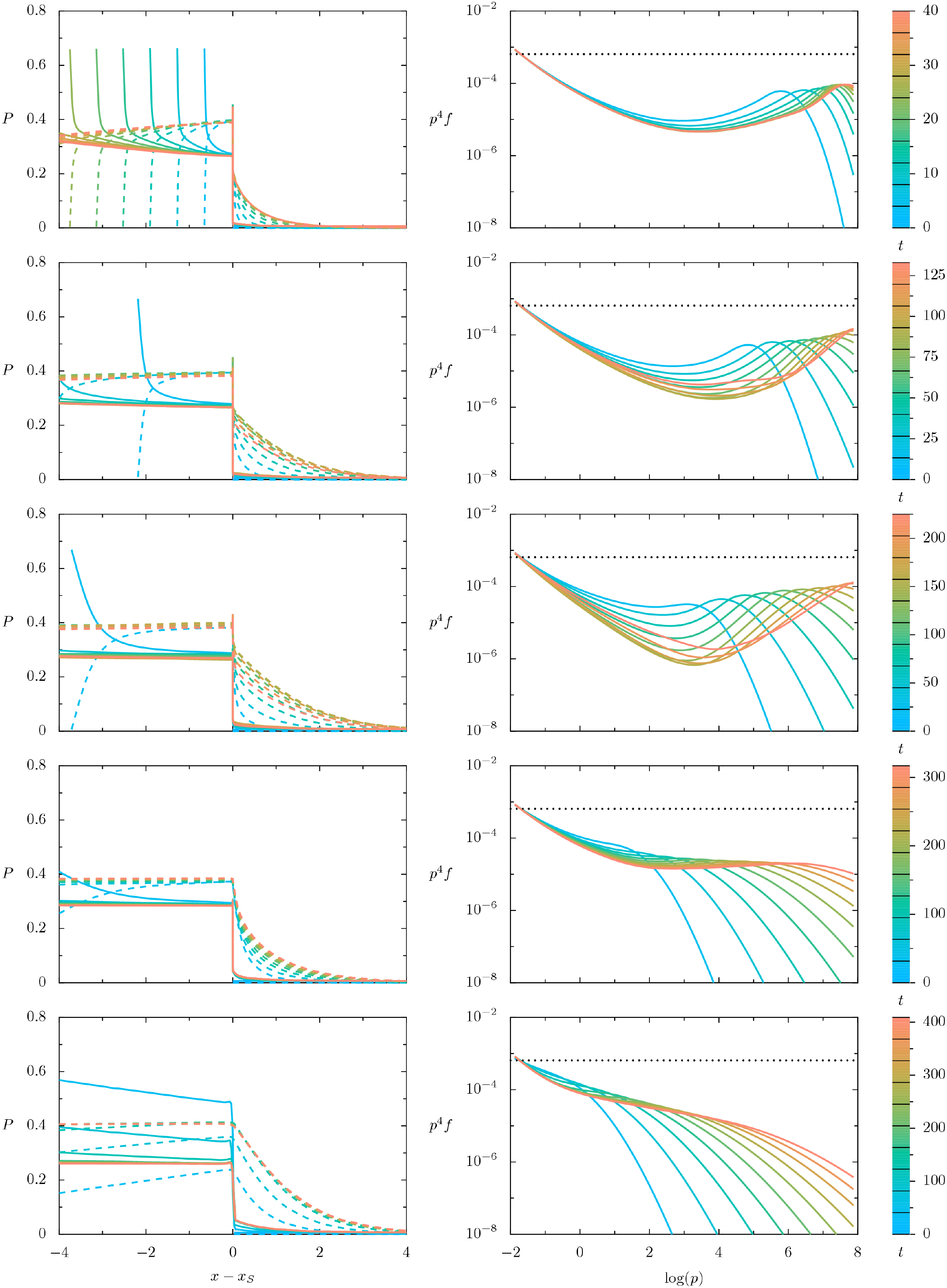}
\caption{Left: Pressure profiles (solid: fluid, dashed: particles) as a function of position at different times.  Right: spectrum of the particles ($g=p^4 f$) as a function of momentum at different times. Diffusion with $\alpha=1/2$ and $p_{\rm c}$ increasing from bottom to top: $10^0$ $m_pc$, $10^2$ $m_pc$, $10^4$ $m_pc$, $10^6$ $m_pc$, and $>10^8$ $m_pc$ (the reference Bohm case). Injection level $\eta=10^{-3}$.}
\label{fig:T6P12_hydro_kino_1e-3}
\end{figure}

We complement our study of the properties of our perpendicular diffusion model with the alternative value $\alpha=1/2$. The simulation results are presented in Figure~\ref{fig:T6P12_hydro_kino_1e-4} for a low injection fraction $\eta=10^{-4}$ and Figure~\ref{fig:T6P12_hydro_kino_1e-3} for a high injection fraction $\eta=10^{-3}$. In this case the values of $p_{\rm c}$, increasing from bottom to top, are $10^0$, $10^2$, $10^4$, $10^6$ $m_p c$, and the topmost plots again correspond to the reference Bohm case, formally recovered when $p_{\rm c} > p_{\rm max} = 10^8\:m_p c$. All the plots presented are otherwise the same as previously.

Both the hydro plots and the kinetic plots are similar to those for $\alpha=1$. Pressure plots all show a modified shock, with an extended precursor, that varies slightly in shape with~$p_{\rm c}$ (and of course with $\eta$). Spectra show a variety of shapes, that evolve similarly as in the previous section with~$p_{\rm c}$. For the reference case of Bohm diffusion (topmost row), there is a clear non-linear curvature, even more marked than for $\alpha=1$. For the modified coefficient, as $p_{\rm c}$ decreases (from top to bottom rows) the curvature is removed, entirely for $\eta=10^{-4}$, and almost entirely for $\eta=10^{-3}$. So for a given $\eta$, the effect is even more marked than for $\alpha=1$, leading to even steeper spectra at low~$p_{\rm c}$. 

\subsection{Discussion}
\label{sec:num:discussion}

\begin{figure}
\includegraphics[scale=1]{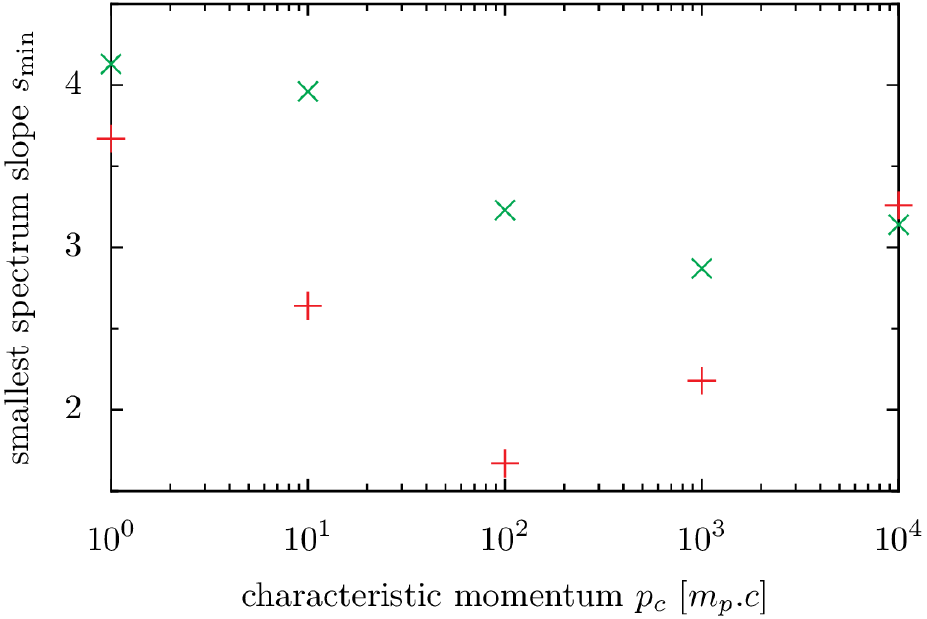}\\
\includegraphics[scale=1]{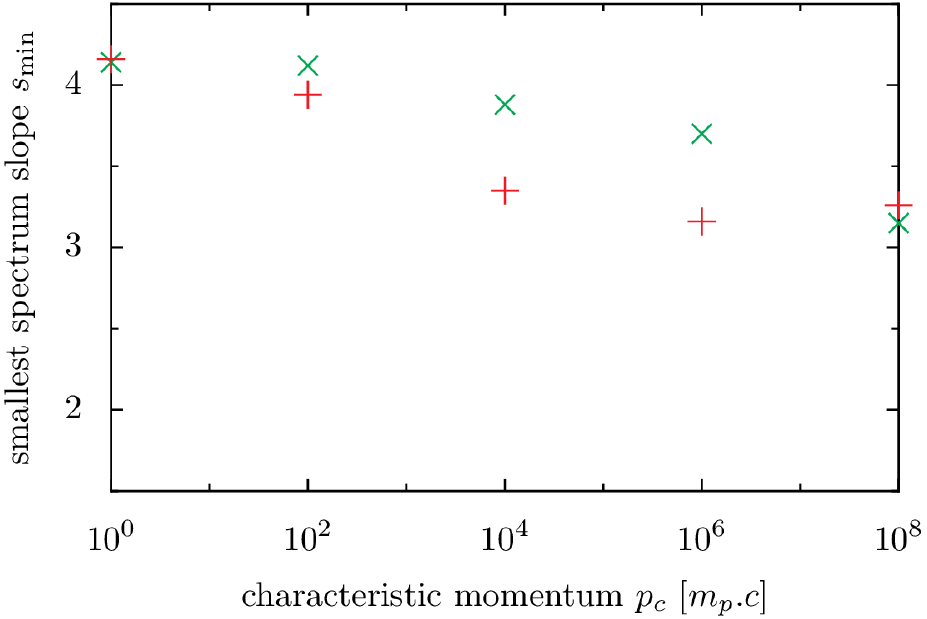}
\caption{Smallest value of the logarithmic slope of the CR distribution function $s=-\partial \ln f / \partial \ln p$, as a function of~$p_{\rm c}$, for different injection levels: $\eta=10^{-4}$ (green crosses) and $\eta=10^{-3}$ (red pluses). The two panels correspond to two different indices for the power-law part of the diffusion law: $\alpha=1$ (top) and $\alpha=1/2$ (bottom).}
\label{fig:smin}
\end{figure}

A~key feature that we have observed in the results of our simulations, for both cases $\alpha=1$ and $\alpha=1/2$, is that our new diffusion coefficient (Equation~(\ref{eq:kNB(p)})) produces modified shocks that clearly bear one distinctive signature of non-linear DSA: the precursor, yet produces spectra that do not bear another landmark signature of non-linear DSA: the concavity. For a very high~$p_{\rm c}$, which is equivalent to the Bohm case, the spectra are fairly hard before $p_{\rm max}$. But as the characteristic momentum~$p_{\rm c}$ is reduced, so is the spectrum curvature. And for a very low~$p_{\rm c}$, the spectra are very steep. To better characterize the transition between the two regimes, in Figure \ref{fig:smin} we plot the minimal value, $s_{\rm min}$, reached by the logarithmic slope of the distribution: $s=-\partial \ln f / \partial \ln p$, as a function of $p_{\rm c}$, for each case of the power-law index ($\alpha=1$ at the top, $\alpha=1/2$ at the bottom) and of the injection fraction ($\eta=10^{-4}$ as green crosses, $\eta=10^{-3}$ as red pluses). We see that $s_{\rm min}$ is close to~3 for the standard case (corresponding to the highest~$p_{\rm c}$), but gets back towards the test-particle value of~4 for the modified case with a diffusion plateau (lowest~$p_{\rm c}$). We also see that the evolution of $s_{\rm min}$ for intermediate cases depends on~$\alpha$ as well as $\eta$. As expected spectra are always harder for a larger injection level, and they react differently to $\eta$ depending on $\alpha$. For the weak momentum dependence $\alpha=1/2$, as well as for the strong momentum dependence $\alpha=1$ with low injection fraction $\eta=10^{-4}$, the variation of $s_{\rm min}$ is regular. For the strong momentum dependence $\alpha=1$ with high injection fraction $\eta=10^{-3}$, the variation of $s_{\rm min}$ is more complicated, with a minimum obtained for intermediate values of~$p_{\rm c}$. Indeed, as can be seen in Figure~\ref{fig:T6P1_hydro_kino_1e-3}, as $p_{\rm c}$ is lowered, the spectrum first sees its concavity being restricted to a smaller range of momenta while still reaching similar values at $p_{\rm max}$, leading to highly modified spectra for that particular parameter range.

We note that the shape of the CR spectrum will be reflected in the shape of the spectrum of non-thermal photons (radiated via the production and decay of pions for energetic protons), so that these findings have direct observational implications. Even though the shock precursor itself has not been resolved, we note that evidence has accumulated indicating that the structure of SNR shocks is indeed affected by the presence of energetic particles \citep{2012SSRv..173..369H}. On the other hand, recent $\gamma$-ray observations of SNRs do not support the existence of concave spectra of the underlying CR population, on the contrary they require rather steep spectra \citep{2011JCAP...05..026C}. We note that our model naturally produces such spectra, even though particles may be efficiently accelerated and take a large part of the shock energy. 

We do not make a claim here that our new diffusion coefficient solves the problem of DSA in SNRs. We recall that this coefficient has been motivated by studies of perpendicular diffusion, but the actual geometry of the magnetic field at SNR shocks is generally uncertain. In principle, both parallel and perpendicular diffusion could be at work. Furthermore, we note that several models have already been proposed, amending the classic theory of non-linear DSA, to accommodate the new data: \cite{2011JCAP...05..026C,2012JCAP...07..038C} stressed the importance of the Alfv{\'e}nic drift, \cite{2012PhPl...19h2901M} and \cite{2012ApJ...755..121B} pointed out the role of neutrals, \cite{2010ApJ...723L.108I} investigated the effect of multiple secondary shocks, and \cite{2012APh....35..300T} showed the possible contribution from the reverse shock. We would like to make the point, however, that some predictions of non-linear DSA regarding the spectrum of particles strongly depend on the diffusion coefficient, the expression of which has often just been guessed or chosen for convenience in studies of particle acceleration in SNRs. We have shown that altering the momentum dependence of~$\kappa$ opens new possibilities. To find out what the real coefficient is, one would need to compute it directly from the magnetic turbulence spectrum, self-consistently with the transport and particles and the development of streaming instabilities. Such a task is well beyond the scope of this paper, where we simply wanted to explore the properties of a new, well motivated form of the diffusion coefficient.


\section{Conclusion}
\label{sec:conclusion}

A key input parameter in studies of particle acceleration in SNRs is the diffusion coefficient in the direction of the shock propagation. It is commonly assumed to follow the Bohm limit, or to be a power-law in momentum with an index~$\alpha$ close to one. In this paper we propose a new model for the diffusion coefficient, based on the work by \cite{2010Ap&SS.325...99S} and \citet{2012AdSpR..49.1643Q} on perpendicular diffusion. Our coefficient reduces to a power-law in momentum for $p<p_{\rm c}$, but becomes independent of the particle momentum for $p>p_{\rm c}$, where $p_{\rm c}$ is some characteristic momentum. 

We provided analytical expressions, in the test-particle regime, to evaluate the maximum momentum $p_{\rm max}$ that can be reached at a given time with this coefficient, which depends on~$p_{\rm c}$ and on the factor~$\kappa_{\infty}$. The two parameters depend on turbulence parameters such as magnetic field strengths and correlation lengths (see \citealt{2010Ap&SS.325...99S} and \citealt{2013ApJ...774....7S}). Such turbulence properties, however, are unknown in SNRs, and so the values of $p_{\rm c}$ and $\kappa_{\infty}$ are poorly constrained and therefore were considered as free parameters. We then performed time-dependent numerical simulations of DSA, including the back-reaction of the particle pressure, with two standard diffusion coefficients (power-laws of index $\alpha=1$ and $\alpha=1/2$) and their modified counterpart (with a varying $p_{\rm c}$ between $1~m_p c$ and $p_{\rm max}(t_{\rm end})=10^4\;m_p c$ or $10^8\;m_p c$). We observed that the shock modifications are similar in all cases, whereas the particle spectra can differ markedly. As long as $p_{\rm max}(t) < p_{\rm c}$ and the diffusion lengths of particles of different~$p$ are well separated, the spectra exhibit the well-known concavity of non-linear DSA. But as soon as $p_{\rm max}(t) > p_{\rm c}$ and the diffusive behaviour of particles becomes independent of their momentum, the spectra become softer. This is in agreement with recent $\gamma$-ray observations of SNRs, which exhibit steep photon spectra, contrary to standard non-linear DSA predictions (see e.g. \citealt{2011JCAP...05..026C} and references therein).

Finally, we note that the applicability of our new model will depend on the shock geometry (the orientation of the ambient magnetic field with respect to the local shock normal). In this paper we made the distinction between fully parallel and fully perpendicular shocks, in reality one would probably need to consider a combination of orientations. In the future we shall study DSA in a multi-dimensional geometry, where all the components of the diffusion tensor may be involved.


\section*{Acknowledgements}
GF, RJD, AS, SSH acknowledge partial support by the Canadian Institute for Theoretical
Astrophysics through the National Fellowships program. 
This work has been also supported by the Natural Sciences and Engineering Research Council of Canada (NSERC) through the Canada Research Chairs and Canada Foundation for Innovation (CFI) funds to SSH, and through a Discovery Grant to AS.
The simulations were carried in large part on a local computing cluster funded by CFI and the Manitoba Research for Innovation Funds (MRIF).
The authors would like to thank Tom Jones for useful discussions, and the anonymous referee for useful comments.

\bibliographystyle{apj}
\bibliography{nonbohm}

\newpage


\end{document}